\documentclass[lettersize,journal]{IEEEtran}
\usepackage{amsmath,amsfonts}
\usepackage{algorithmic}
\usepackage{algorithm}
\usepackage{array}
\usepackage{textcomp}
\usepackage{stfloats}
\usepackage{url}
\usepackage{verbatim}
\usepackage{graphicx}
\usepackage{cite}
\usepackage{bm}
\usepackage{tikz}
\usepackage{comment}
\usepackage{siunitx}
\usepackage[subrefformat=parens]{subcaption}
\usepackage{stfloats}

\usepackage{CJKutf8}

\makeatletter
\def\ps@IEEEtitlepagestyle{%
  \def\@oddhead{}%
  \def\@evenhead{}%
  \def\@oddfoot{\hfil\footnotesize
  This work has been submitted to the IEEE for possible publication. Copyright may be transferred without notice, after which this version may no longer be accessible.\hfil}%
  \def\@evenfoot{\@oddfoot}%
}
\makeatother

\begin{document}

\begin{CJK*}{UTF8}{ipxm}

\title{Variable-Impedance Muscle Coordination \\ under Slow-Rate Control Frequencies \\and Limited Observation Conditions \\ Evaluated through Legged Locomotion}

\author{
Hidaka Asai, Tomoyuki Noda,~\IEEEmembership{Member, IEEE} and Jun Morimoto~\IEEEmembership{Member, IEEE}
\thanks{Hidaka~Asai, Tomoyuki Noda, and Jun Morimoto are with the Department of Brain Robot Interface, ATR Computational Neuroscience Laboratories, Kyoto, 619-0288, Japan; E-mail: (hidaka.asai@atr.jp (HA), 
t\_noda@atr.jp (TN),
xmorimo@atr.jp (JM)), Hidaka~Asai and Jun Morimoto are also with the Kyoto University Graduate school of informatics.}
\thanks{
This work was supported by the project , JAPAN,  Japan Science and Technology Agency (JST) Moonshot R\&D, Grant Number: JPMJMS2034, JSPS KAKENHI Grant Number 24K21325, and the Tateishi Science and Technology Foundation, JPNP20006, commissioned by NEDO.
}
}

\markboth{Journal of \LaTeX\ Class Files,~Vol.~14, No.~8, August~2021}%
{Shell \MakeLowercase{\textit{et al.}}: A Sample Article Using IEEEtran.cls for IEEE Journals}



\maketitle

\begin{abstract}
Human motor control remains agile and robust despite limited sensory information for feedback,
a property attributed to the body’s ability to perform morphological computation through muscle coordination with variable impedance.
However, it remains unclear how such low-level mechanical computation reduces the control requirements of the high-level controller.
In this study, we implement a hierarchical controller consisting of a high-level neural network trained by reinforcement learning
and a low-level variable-impedance muscle coordination model with mono- and biarticular muscles
in monoped locomotion task.
We systematically restrict the high-level controller by varying the control frequency 
and by introducing biologically inspired observation conditions: delayed, partial, and substituted observation.
Under these conditions, we evaluate how the low-level variable-impedance muscle coordination contributes to 
learning process of high-level neural network.
The results show that variable-impedance muscle coordination enables stable locomotion even 
under slow-rate control frequency and limited observation conditions.
These findings demonstrate that the morphological computation of muscle coordination 
effectively offloads high-frequency feedback of the high-level controller
and provide a design principle for the controller in motor control.
\end{abstract}


\section{Introduction}
\label{sec:Introduction}

\begin{figure}[tb]
            \centering
	    \includegraphics[width=\hsize]{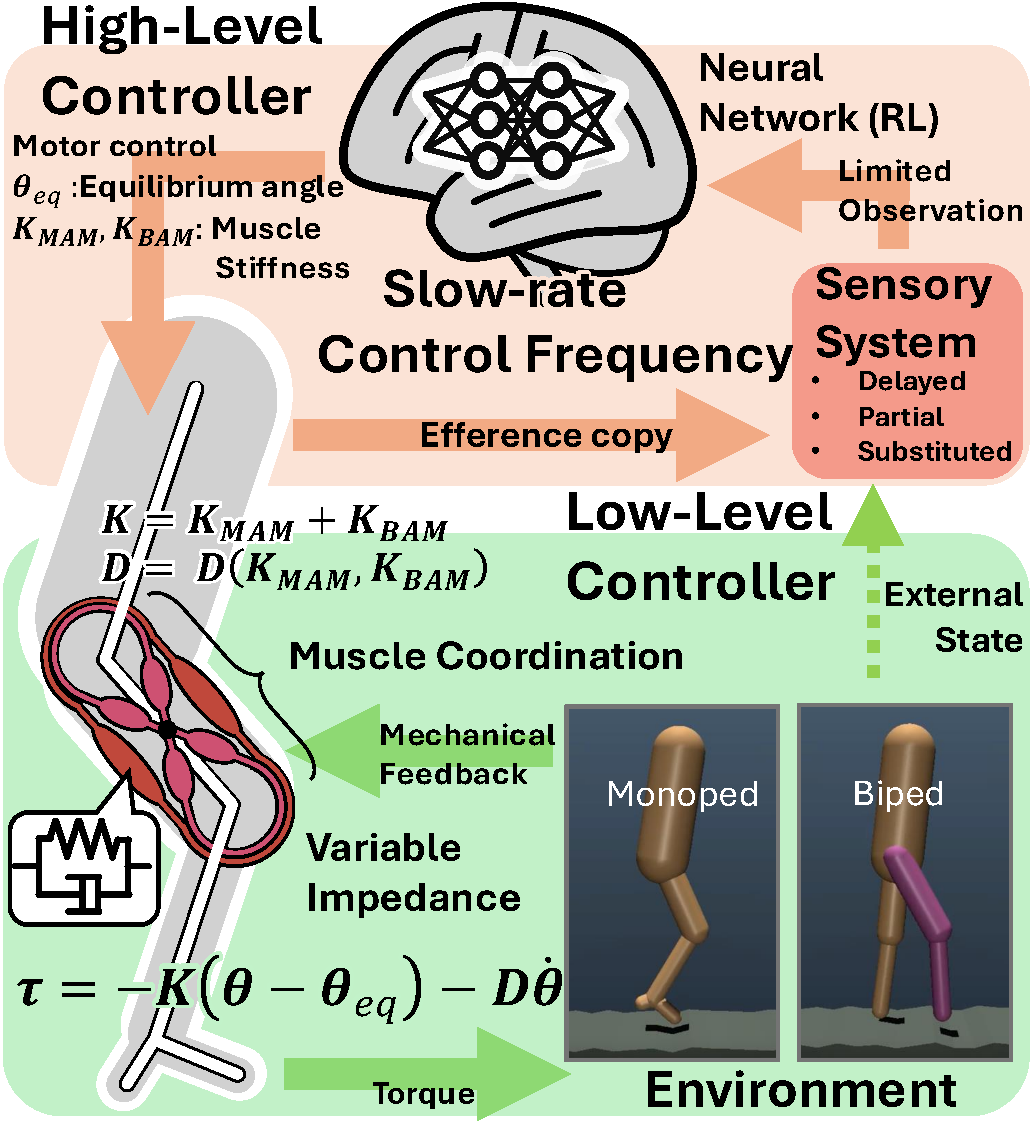}
            \caption{
            Proposed bio-inspired hierarchical control architecture.
            The high-level neural-network controller operates at a slow-rate control frequency 
            with limited observations,
            while the low-level controller, 
            implemented by variable-impedance muscle coordination, 
            generates torques through mechanical feedback.
            Morphological computation in the low-level controller simplifies the control requirement of the high-level controller.
            }
            \label{Fig:hierachy_control_diagram}
\end{figure}

Physiological motor control is remarkably agile and robust despite delayed and partial observations \cite{MIALL19961265}.
One contributing factor is that humans possess a mechanical controller realized by coordinated muscles with
variable impedance\cite{Burdet2001_OptimalImpedance}.
Coordinated activations of mono- and biarticular muscles modulate joint impedance and allow the system to cope with fast dynamics through the body’s own mechanical feedback, without relying on high-bandwidth neural feedback \cite{hogan1985impedance}.
Moreover, musculoskeletal impedance is known to change dynamically during movement \cite{gomi1998task}.
In this way, the variable impedance of muscle coordination enables the body to handle a portion of the control effort mechanically, 
which is a representative form of morphological computation \cite{pfeifer2006body}.




From this perspective, human motor control can be understood as a hierarchical control system composed of
brain and nervous system as high-level controller and muscle coordination as low-level controller.
The low-level controller provides mechanical feedback 
through body structures, including morphological computation through the musculoskeletal system, which is often referred to as embodied intelligence \cite{brooks1991intelligence} or physical intelligence \cite{SITTI2021101340}.
Musculoskeletal robots driven by pneumatic artificial muscles have demonstrated 
that mimicking human muscle arrangements and coordination patterns enables stable motion generation 
without sophisticated feedback control \cite{iida2008bipedal, takuma20083d, niiyama2010athlete},
supporting the view that muscle coordination contributes to embodied intelligence.
In complex systems where high-level feedback is limited, 
ordered dynamics can emerge through interaction between a well-designed low-level controller and environment.
This process is referred to as "self-organization" \cite{pfeifer2007self},
indicating that part of the computation required for stable motor control
is handled by the low-level body mechanics itself rather than solely by high-level feedback.
This low-level information processing by morphology is termed morphological computation\cite{pfeifer2006body}.



\begin{figure*}[tb]
	\centering
    \begin{minipage}{0.3\hsize}
	    \centering
	    \subfloat[Fixed impedance (FI)]{\includegraphics[width=0.9\hsize]{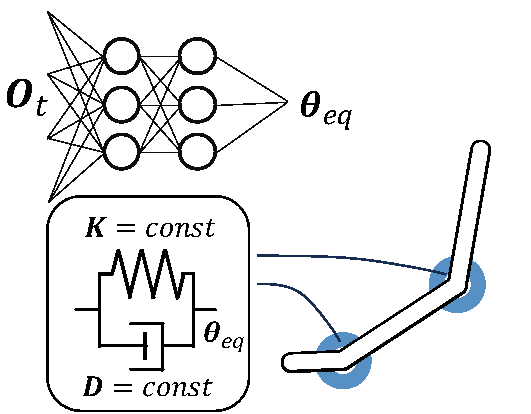}}
	\end{minipage}
	\begin{minipage}{0.3\hsize}
	    \centering
	    \subfloat[Monoaticular muscle (MO)]{\includegraphics[width=0.9\hsize]{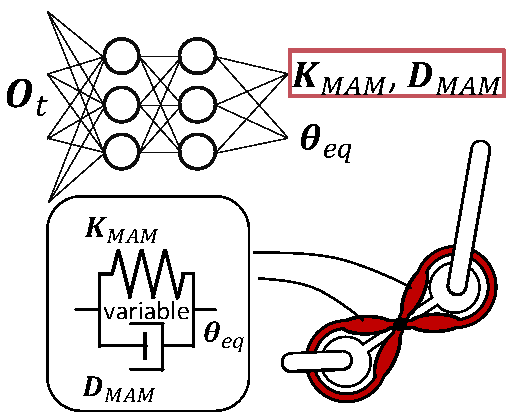}}
	\end{minipage}
        \begin{minipage}{0.3\hsize}
	    \centering
	    \subfloat[Muscle coordination (MC)]{\includegraphics[width=0.9\hsize]{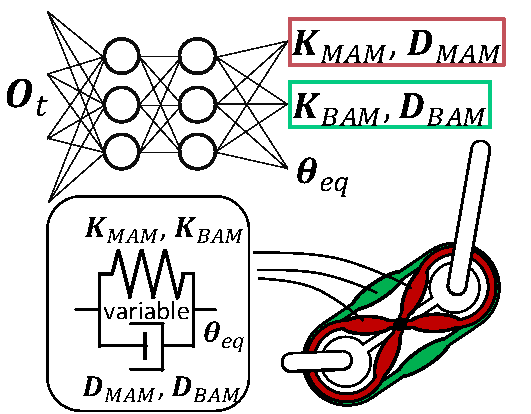}}
	\end{minipage}
	\caption{
     Comparison of controller architectures:
    (a) Fixed impedance (FI): joint gains are fixed and do not change during learning.
    (b) Monoarticular variable impedance (MO): each joint stiffness is varied at every time step.
    (c) Variable impedance muscle coordination (MC): in addition to joint-wise gains, biarticular impedance spanning two joints is varied.
    }
        \label{fig:controllers}
\end{figure*}

When the morphology of the low-level controller changes, the information-processing demands on the high-level controller also change \cite{lungarella2006mapping}.
In this sense, morphology determines sensor–motor coordination \cite{pfeifer2007self} and the requirements imposed on the high-level controller \cite{paul2006morphological}.
Thus, when a well-designed morphology functions as a low-level controller, morphological computation can simplify 
the requirements of the high-level controller.



The ability of muscles and muscle coordination to perform morphological computation
has been analyzed quantitatively in terms of mechanically processed information in \cite{ghazi2016evaluating}. 
In this context, Neural networks have been employed as high-level controllers\cite{haeufle2020muscles}.
Furthermore, incorporating variable impedance into the low-level controller has been shown 
to improve robustness in contact-rich tasks such 
as hopping and wiping \cite{martin2019variable, zhang2024srl, bogdanovic2020learning}.
However, in conventional studies, it remains unclear how such low-level morphological computation 
determines the requirements of high-level controller and
contributes to the design of the high-level controller. 
It is unclear under what conditions variable impedance is advantageous and under what conditions it is disadvantageous.
Especially with variable impedance, the increased dimensionality of the action should create some conditions that slow down the learning of neural networks.
This reason makes it difficult to apply the morphological computation to engineering use.


In this study, we aim to clarify how variable-impedance muscle coordination at the low-level controller contributes to the design requirements imposed on the high-level controller.
Based on 
\cite{pfeifer2007self}, we implement a bio-inspired hierarchical controller illustrated in Fig.~\ref{Fig:hierachy_control_diagram}. 
As the target task, monoped and biped locomotion, mainly monoped locomotion, in the sagittal-plane is considered.
The low-level controller is modeled as muscle coordination including biarticular muscles with variable impedance
and each muscle pair is modeled as a combination of a variable stiffness and a variable damper.
The high-level controller is modeled as neural network trained via reinforcement learning with the low-level controller.
We deliberately constrained the ability of the high-level controller by imposing a slow-rate control frequency and 
limited observation conditions, 
delayed (representing transmission delay), 
partial (lack of angular velocity information), and substituted 
(lack of actual joint angle information, replaced by an efference copy) observations.
Under these conditions, we compare low-level muscle-coordination controller with other controllers 
to assess how the morphological computation changes the performance demands on the high-level controller.

The contributions of this paper are summarized as follows:
\begin{quote}
    \begin{itemize}
        \item 
        Variable impedance in the low-level controller enables 
        stable legged locomotion even at a slow-rate control frequency (3 Hz)
        \item 
        Variable-impedance muscle coordination allows locomotion under the limited sensory feedback conditions. 
        \item 
        Variable-impedance muscle coordination achieves monoped locomotion 
        using only efference copy instead of actual leg angle. 
    \end{itemize}
\end{quote}

The remainder of this paper is organized as follows.
Section \ref{sec:Related_works} reviews related works about the morphological computation of variable-impedance muscle coordination.
Section \ref{sec:Model} formulates the variable-impedance muscle coordination controller and observation conditions in detail.
The simulation results comparisons with the other controllers are presented in Section \ref{sec:Experiment}, 
and their implications are discussed in Section \ref{sec:Discussion}.
Finally, Section \ref{sec:conclusion} summarizes the paper.

\section{Related works}
\label{sec:Related_works}

\subsection{Muscle coordination}
\label{sec:related_mc}

Human motion control involves a large number of degrees of freedom, and accurately describing body dynamics requires a high-dimensional representation,
which poses a major challenge for control \cite{bernstein1967co}.
There are also delays in neural feedback from observed states \cite{WOLPERT1998338}.
Consequently, designing inputs instantaneously from all states at a fast-rate control frequency is unrealistic.
One mechanism to cope with these limitations is muscle coordination, which performs morphological computation to stabilize motion.
In motion generation as well, muscle coordination is argued to contribute to morphological computation \cite{pfeifer2006body}.
In robotics, mimicking human muscle arrangements and activation patterns enables robust walking and running without precise feedback control \cite{iida2008bipedal, takuma20083d, niiyama2010athlete}.
In particular, muscle coordination including biarticular muscles can control ground reaction 
force direction \cite{hof2001force, kaneko2016force} and improve dynamical stability in running \cite{asai2025dynamical},
These studies support the interpretation that muscle coordination performs morphological computation in the system dynamics.

 \subsection{Morphological computation}

Classic examples of embodied intelligence include passive dynamic walking \cite{mcgeer1990passive} and the spring-loaded inverted pendulum (SLIP) model \cite{blickhan1989spring},
which generate self-organized and self-stable motion without high-level feedback control.
These works indicate that a well-designed morphology can make an explicit high-level controller less necessary.
Such low-level information processing by body structure is termed morphological computation, 
and morphology is known to shape sensor–motor relation
and influence high-level control requirements \cite{pfeifer2007self, lungarella2006mapping, paul2006morphological}.
However, how the morphological computation of muscle coordination changes the performance 
required of the high-level controller has not been clearly analyzed.

\subsection{Variable impedance and reinforcement learning}


Variable impedance, as observed in human muscle coordination, 
is considered crucial for robots operating in contact-rich tasks \cite{liu2025human}.
Approaches that combine a variable-impedance low-level controller with a reinforcement-learned high-level neural network
have demonstrated higher robustness than fixed-impedance baselines 
in contact-rich tasks, such as hopping and wiping \cite{martin2019variable, zhang2024srl, bogdanovic2020learning}.
However, these studies primarily focus on task performance, and do not systematically analyze 
how variable impedance modifies the performance requirements of the high-level controller.
Even though variable impedance has benefits in morphological computation,
it causes an increase in dimensionality of the neural network search parameter space due to the addition of action dimensions, and
there should be a trade-off between these advantages and disadvantages.
To apply the morphological computation to engineering use,
this trade-off must be clarified.




Building on these lines of work, this study integrates muscle coordination, 
variable impedance, and reinforcement learning to examine
how the morphological computation provided by variable-impedance muscle coordination 
changes the sensor-motor relation and
simplifies the requirements of a high-level neural-network controller.
We systematically restrict the high-level controller in terms of control bandwidth (slow-rate control frequency) 
and limited observation conditions, 
and compare how far the lower-level morphological computation compensates for these constraints.
To this end, we compare three types of lower controllers (Fig.~\ref{fig:controllers}):
(a) Fixed impedance controller (FI),
(b)  variable-impedance monoarticular-muslce controller (MO), and
(c) variable-impedance muscle-coordination controller with biarticular muscle (MC).
This comparison reveals the impact of the morphological computation 
of the mucle coordination
on the performance requirements of the high-level controller.

\section{Model}
\label{sec:Model}



In this study, we focus on the monoped and biped locomotion task, mainly monoped locomotion, as a benchmark  
to evaluate the effects of variable-impedance muscle coordination on hierarchical control performance
in dynamic motion control.

\subsection{Muscle coordination as low-level controller}

\begin{figure}[tb]
	\centering
	\begin{tabular}{cc}
        \begin{minipage}{0.4\hsize}
	    \centering
	    \includegraphics[height=45mm]{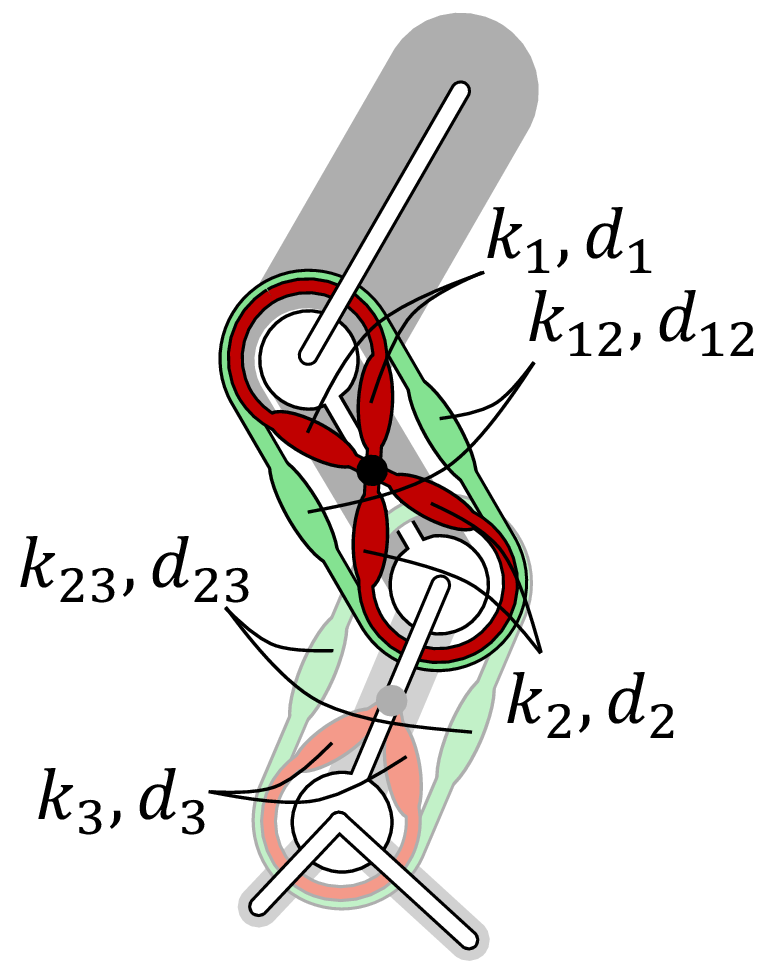}
            \subcaption{Muscle setting }
            \label{fig:muscle_coordination_monoped}
	\end{minipage}
        \begin{minipage}{0.4\hsize}
	    \centering
	    \includegraphics[height=45mm]{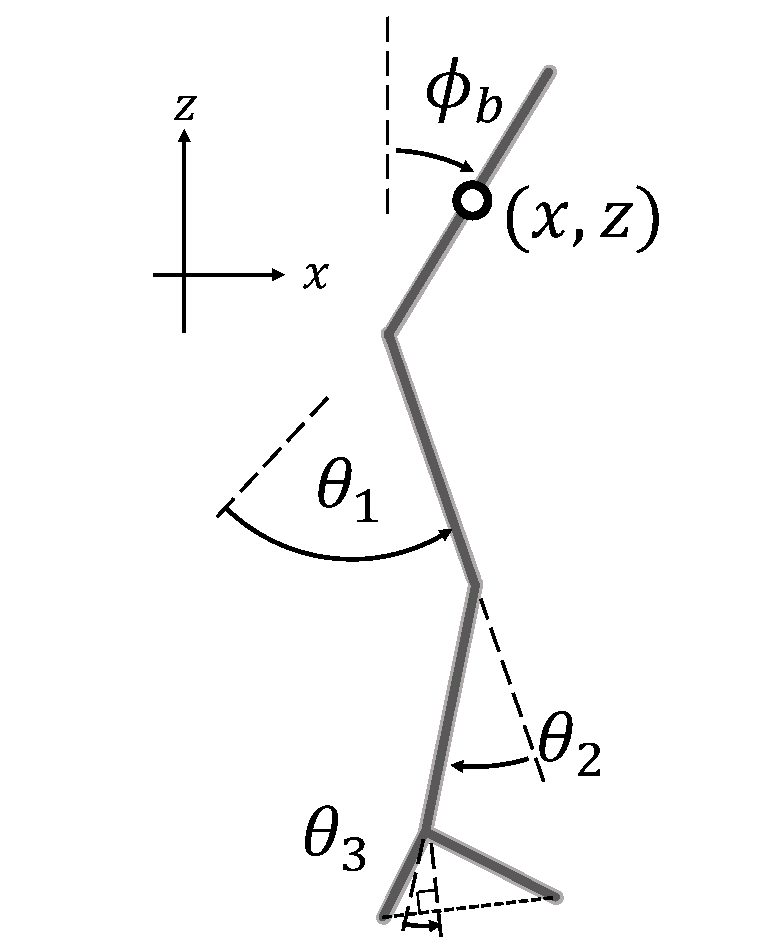}
            \subcaption{Coordinate setting }
            \label{fig:model_coordinate_monoped}
	\end{minipage}
	\end{tabular}
	\caption{Monoped locomotion model.
    The model consists of a torso, thigh, shank, and foot.}
\end{figure}

 \begin{figure}[tb]
 	\centering
 	\begin{tabular}{cc}
     	\begin{minipage}{0.4\hsize}
 	    \centering
 	    \includegraphics[height=45mm]{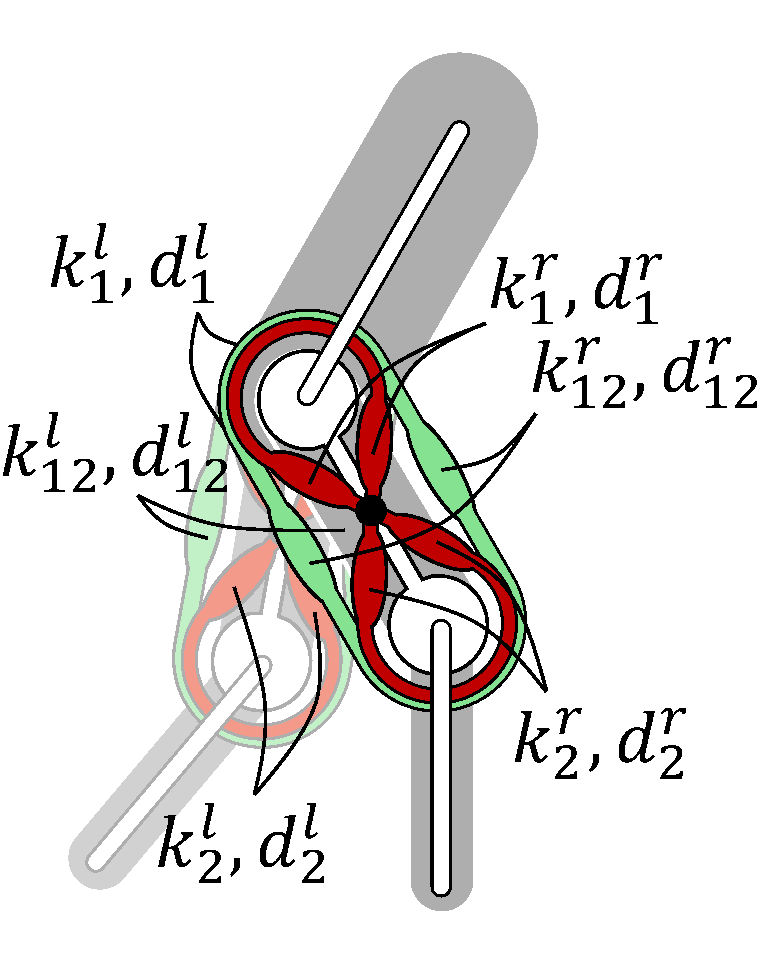}
             \subcaption{Muscle setting}
             \label{fig:muscle_coordination_biped}
 	\end{minipage}
         \begin{minipage}{0.4\hsize}
 	    \centering
 	    \includegraphics[height=45mm]{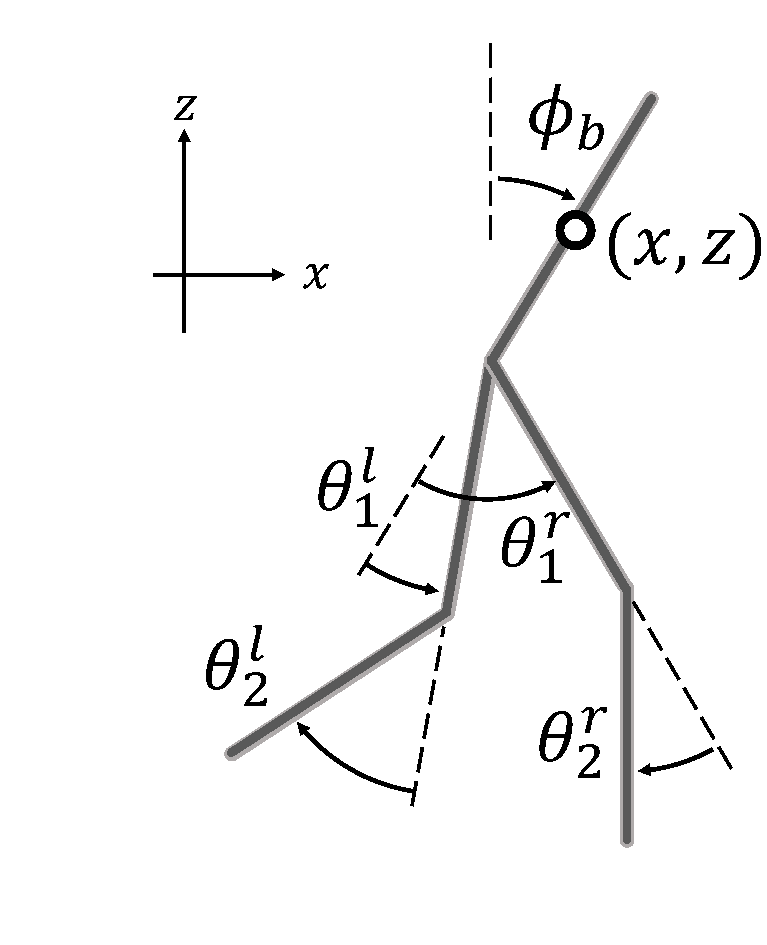}
             \subcaption{Coordinate setting}
             \label{fig:model_coordinate_biped}
 	\end{minipage}
 	\end{tabular}
 	\caption{Biped locomotion model.
    The model consists of a torso, two thighs, and two shanks with no foot.}
 \end{figure}

Muscle coordination of a sagittally constrained human being is modeled 
by monoarticular and biarticular muscles.
The muscles of the monoped locomotion model is shown in Fig. \ref{fig:muscle_coordination_monoped} and
that of the biped lcoomotion model is shown in Fig. \ref{fig:muscle_coordination_biped}.
Monoarticular muscles actuate one joint, 
while a biarticular muscle actuates two joints,
represented by red and green in the figures.
The torque, generated by the muscles around joint is described as
\begin{align}
    \bm{\tau} &= - \bm{K} \left( \bm{\theta} -\bm{\theta}_{eq} \right) - \bm{D}  \dot{\bm{\theta}.}    \label{eq:torque_lim}
\end{align}
$\bm{\tau}$, $\bm{\theta}$ and $\bm{\theta}_{eq}$ are the vector of joint torques, joint angles and equilibrium joint angles.
$\bm{K}$ and $\bm{D}$ are the stiffness matrix and the damping matrix,
designed by the muscle coordination.
A pair of muscles is represented by a single linear torsion stiffness and a single torsion damping coefficient.
$\bm{K}$ and $\bm{D}$ are expressed as the addition of 
monoarticular muscle component $\bm{K}_{MAM}$, $\bm{D}_{MAM}$ and 
biarticular muscle component $\bm{K}_{BAM}$, $\bm{D}_{BAM}$ as
\begin{align}
    \bm{K} = \bm{K}_{MAM} + \bm{K}_{BAM}\\
    \bm{D} = \bm{D}_{MAM} + \bm{D}_{BAM},
\end{align}
where MAM and BAM stand for Mono-Articular Muscle and Bi-Articular Muscle, respectively.

In the monoped locomotion task, as shown Fig. \ref{fig:model_coordinate_monoped},
the leg consists of a thigh, a shank, and a foot and
the model has a hip joint, a knee joint, and an ankle joint, 
accordingly.
$\bm{\tau}$, $\bm{\theta}$ and $\bm{\theta}_{eq}$ are defined as
$\bm{\tau} = \left[ \tau_{1}, \tau_{2}, \tau_{3} \right]^T$, 
$\bm{\theta} = \left[ \theta_{1}, \theta_{2}, \theta_{3} \right]^T$ and
$\bm{\theta}_{eq} = \left[ {\theta_{eq1}}, {\theta_{eq2}}, {\theta_{eq3}} \right]^T$ with
the indexes $1$, $2$ and $3$ meaning the variables about hip, knee and ankle joint, respectively.
$\bm{K}_{MAM}$, $\bm{D}_{MAM}$
are defined as
\begin{align}
    \bm{K}_{MAM} &= 
    \left[
    \begin{array}{ccc}
         k_1 & 0   & 0  \\
         0   & k_2 & 0  \\
         0   & 0   & k_3
    \end{array}
    \right]\\
    \bm{D}_{MAM} &= 
    \left[
    \begin{array}{ccc}
         d_1 & 0   & 0  \\
         0   & d_2 & 0  \\
         0   & 0   & d_3
    \end{array}
    \right].
\end{align}
Assuming that the lever arm of the biarticular muscle is the same at each joint,
$\bm{K}_{BAM}$, and $\bm{D}_{BAM}$ are defined as
\begin{align}
    \bm{K}_{BAM} &= 
    \left[
    \begin{array}{ccc}
         k_{12} & k_{12}   & 0  \\
        k_{12} & k_{12} + k_{23}   & k_{23}  \\
         0      & k_{23}   & k_{23}
    \end{array}
    \right]\\
    \bm{D}_{BAM} &= 
    \left[
    \begin{array}{ccc}
         d_{12} & d_{12}   & 0  \\
        d_{12} & d_{12} + d_{23}   & d_{23}  \\
         0      & d_{23}   & d_{23}
    \end{array}
    \right].
\end{align}
$k$ and $d$ express the stiffness coefficient and damping coefficient of each muscle pair.
Primates, including humans, do not have anterior lower biarticular muscle, 
but that is not considered in this study.

In the biped locomotion task, as shown Fig. \ref{fig:model_coordinate_biped},
the model does not have feet and ankle joints,
to ensure uniformity in ways of walking through the results of the learning.
In this simulation setting,
the model has two hip joints and two knee joints.
$\bm{\tau}$, $\bm{\theta}$ and $\bm{\theta}_{eq}$ are defined as
$\bm{\tau} = \left[ \tau_{1}^l, \tau_{2}^l, \tau_{1}^r, \tau_{2}^r \right]^T$, 
$\bm{\theta} = \left[ \theta_{1}^l, \theta_{2}^r,\theta_{1}^r, \theta_{2}^r \right]^T$ and
$\bm{\theta}_{eq} = \left[ {\theta_{eq1}}^l, {\theta_{eq2}}^l, {\theta_{eq1}}^r, {\theta_{eq2}}^r \right]^T$.
Similar to monoped locomotion model,
$\bm{K}_{MAM}$, $\bm{D}_{MAM}$, $\bm{K}_{BAM}$ and $\bm{D}_{BAM}$ are described as
\begin{align}
    \bm{K}_{MAM} &= 
    \left[
    \begin{array}{cccc}
         k_1^l & 0     & 0     & 0  \\
         0     & k_2^l & 0     & 0  \\
         0     & 0     & k_1^r & 0  \\
         0     & 0     & 0     & k_2^r 
    \end{array}
    \right]\\
    \bm{D}_{MAM} &= 
    \left[
    \begin{array}{cccc}
         d_1^l & 0     & 0     & 0  \\
         0     & d_2^l & 0     & 0  \\
         0     & 0     & d_1^r & 0  \\
         0     & 0     & 0     & d_2^r 
    \end{array}
    \right]\\
    \bm{K}_{BAM} &= 
    \left[
    \begin{array}{cccc}
        k_{12}^l & k_{12}^l & 0        & 0      \\
        k_{12}^l & k_{12}^l & 0        & 0      \\
        0        & 0        & k_{12}^r & k_{12}^r \\
        0        & 0        & k_{12}^r & k_{12}^r
    \end{array}
    \right]\\
    \bm{D}_{BAM} &= 
    \left[
    \begin{array}{cccc}
        d_{12}^l & d_{12}^l & 0        & 0      \\
        d_{12}^l & d_{12}^l & 0        & 0      \\
        0        & 0        & d_{12}^r & d_{12}^r \\
        0        & 0        & d_{12}^r & d_{12}^r
    \end{array}
    \right].
\end{align}
Index $l$ and $r$ mean the variables about the left leg and the right leg,
respectively.

For both locomotion model,
once the stiffness value of one muscle pair $k$ is determined,
the corresponding damping value
$d$ is automatically determined by the stiffness value as
\begin{align}
    d &= \sqrt{ c_{kd} \cdot k }, \label{eq:k_to_d}
\end{align}
where $c_{kd}$ is a constant coefficient,  
and
\begin{equation}
       \bm{D} = \bm{D}_{MAM} ( \bm{K}_{MAM} ) + \bm{D}_{BAM} ( \bm{K}_{BAM} ). 
\end{equation}

\subsection{Observation conditions}

In monoped locomotion task,
five observation conditions are prepared
to investigate the effect of muscle coordination on observation requirements of high-level controller.

\subsubsection{Nominal observation}
The nominal observation condition is 
set as the same with \textit{Hopper-v4} environment in \textit{Open-AI Gym} \cite{brockman2016openai} as
\begin{align}
    \bm{o}_t = \left[ z, \phi_b, \theta_1, \theta_2, \theta_3, \dot{x}, \dot{z}, \dot{\phi}_b, \dot{\theta}_1, \dot{\theta}_2, \dot{\theta}_3 \right]
    \in \mathbb{R}^{11}.
\end{align}
As shown in Fig. \ref{fig:model_coordinate_monoped},
$x$ is the horizontal position of the center of torso mass,
$z$ is the vertical position of the center of torso mass,
and
$\phi_b$ is the tilt of a torso.
Dots mean the time derivative of each state.
This condition consists of each position and each velocity, except $x$.
In continuous systems,
this state represents the precise description of the equations of motion for the hopper as a 4-link system.

\subsubsection{Delayed observation}

The delayed observation condition consists of the observed state and the action from one time step before:
\begin{align}
    \bm{o}_t &= \left[ z^{t-1}, \phi_{b}^{t-1}, \theta_{1}^{t-1}, \theta_{2}^{t-1}, \theta_{3}^{t-1}, \dot{x}^{t-1}, \dot{z}^{t-1}, \right. \nonumber \\ 
     & \quad \left. \dot{\phi}_{b}^{t-1},  \dot{\theta}_{1}^{t-1}, \dot{\theta}_{2}^{t-1}, \dot{\theta}_{3}^{t-1}, \bm{a}^{t-1} \right]
     \in \mathbb{R}^{11 + d_a } \label{eq:obs_delayed} .
\end{align}
This condition represents the transmission delay commonly observed in biological systems, 
where immediate sensory feedback is not fully available. 
Along with the previous state, the previous action $\bm{a}^{t-1} \in \mathcal{A} \subseteq \mathbb{R}^{d_a}$, 
normalized input to each low-level controller is also included, 
emulating the role of an efference copy. 
To generate appropriate feedback, 
the high-level neural network must implicitly infer the current state from the previous state and motor command memory.
This condition allows us to evaluate whether the low-level mechanical controller’s 
morphological computation produces dynamics that are easier to infer 
and more robust against feedback delay, from the perspective of high-level controller.
The delay duration $T_{delay}$ corresponds to one control time step of the high-lelvel controller, 
and is designed as $T_{delay} = 1/\nu_c$ by using the control frequency $\nu_c$.
Therefore, the delay becomes shorter at a fast-rate control frequency 
and longer at a slow-rate control frequency.


\subsubsection{Partial observation}

The partial observation condition is expressed as
\begin{align}
    \bm{o}_t = \left[ z, \phi_b, \theta_1, \theta_2, \theta_3, \dot{x}, \dot{z}\right]
    \in \mathbb{R}^7,
\end{align}
without the information of each angular velocity.
In this state representation, 
the high-level controller cannot observe angular velocity information, 
then the higher-level neural networks cannot perform feedback control corresponding to angular velocity, fast dynamics.
Therefore, only the low-level mechanical controller must handle fast dynamics
with its morphological computation ability.


\subsubsection{Delayed partial observation}

Delayed partial observation is expressed as
\begin{align} 
    \bm{o}_t =& \left[ z^{t-1}, \phi_{b}^{t-1}, \theta_{1}^{t-1}, \theta_{2}^{t-1}, \theta_{3}^{t-1}, \right. \nonumber \\
    & \quad \left. \dot{x}^{t-1}, \dot{z}^{t-1}, \bm{a}^{t-1} \right]
    \in \mathbb{R}^{7 + d_a } ,
\end{align}
which is (\ref{eq:obs_delayed}) minus joint angular velocities one step before.
This observation condition is the combination of the delayed and partial observation conditions.
It can be measured whether the low-level controller can generate low-level dynamics 
that are easier to compensate the observation delays 
for the high-level controller in the situation that the fast dynamics is unobservable.


\subsubsection{Substituted observation}
The Substituted observation is defined as
\begin{align}
    \bm{o}_t = \left[ z, \phi_b, \bm{a}_{\theta}^{t-1}, \dot{x}, \dot{z}\right]
    \in \mathbb{R}^{7},
\end{align}
$\bm{a}^{t-1}$ contains the normalized equilibrium angles command and, optionally, the normalized stiffness command. 
Among them, $\bm{a}^{t-1}_{\theta}$ denotes the part of $\bm{a}^{t-1}$ corresponding to 
the equilibrium angles as $\bm{\theta}_{eq} = [ \theta_{eq1}, \theta_{eq2}, \theta_{eq3}]^T$.
In this observation condition, the current joint angles ${\bm\theta}$ is not directly observed.
The equilibrium angles commanded one time step before are used as substitutes  as efference copy instead of ${\bm\theta}$.
In fact, humans are unable to sense precise joint angles directly. 
Instead, they estimate body posture and movement by integrating somatosensory feedback
with internal predictions generated from motor commands via the efference copy.
This observation condition mimics that mechanism.
Afferent feedback from the leg is unavailable, and 
only the efference copy of the previous equilibrium command is available.
This design is conceptually related to the equilibrium point hypothesis \cite{bizzi1984posture}, 
in which voluntary motion is generated by shifting the equilibrium positions of the limbs.

Under this condition, we can evaluate whether the mechanical low-level controller 
enables the physical leg to follow the equilibrium trajectories produced by the high-level network, 
and consequently, whether the actual joint-angle sensing can be omitted for achieving task.
Because the locomotion task involves strong ground contacts, 
the task requires producing ground reaction forces in appropriate directions and magnitude. 
Therefore, this condition is expected to highlight the advantage of human-like 
muscle coordination, which naturally couples stiffness and force direction
as described in Section \ref{sec:related_mc}.


In biped locomotion task, only two observation conditions,
the nominal observation and the substituted observation,
are prepared.
This is because,
biped locomotion with sagittal constraints 
is too stable to see the difference between results under various observational conditions unlike monoed locomotion.
The nominal observation is expressed as
 \begin{align}
     \bm{o}_t &= \left[ z, \phi_b, \theta_1^l, \theta_2^l,  \theta_1^r, \theta_2^r, \right. \nonumber \\
     &\qquad\left. \dot{x}, \dot{z}, \dot{\phi}_b, \dot{\theta}_1^l, \dot{\theta}_2^l,  \dot{\theta}_1^r, \dot{\theta}_2^r \right]
     \in \mathbb{R}^{13},
 \end{align}
and
the Substituted observation is expressed as
 \begin{align}
     \bm{o}_t = \left[ z, \phi_b, \bm{a}_{\theta}^{t-1},
     \dot{x}, \dot{z} \right]
     \in \mathbb{R}^{8},
 \end{align}
where $\bm{a}^{t-1}_{\theta}$ represents the normalized $\bm{\theta}_{eq} = [ \theta^l_{eq1}, \theta^l_{eq2},\theta^r_{eq1}, \theta^r_{eq2}]^T$ as a part of the action
 

\section{Experiment} \label{sec:Experiment}

In this study,
we prepare target velocity tracking task at $1\si{m/s}$ for monoped and biped locomotion task.
The learning processes of high-level neural-network controllers 
for each low-level controller are evaluated.

\subsection{Low-level controller and model settings}

In the task, three low-level controllers are compared,
fixed impedance (FI),
variable-impedance monoarticular muscle (MO), and
Variable-impedance muscle-coordination (MC)
as shown in Fig.\ref{fig:controllers}.
The outputs of the neural networks are normalized 
and converted into physically meaningful inputs through linear transformations.
The input to FI provided by the high-level network is $\bm{\theta_{eq}}$,
the input to MO is composed of $\bm{K}_{MAM}$ and $\bm{\theta_{eq}}$,
and
the input to MC is composed of $\bm{K}_{MAM}$, $\bm{K}_{BAM}$, and $\bm{\theta_{eq}}$.
The maximum value of each variable stiffness is designed as $1500 ~ \si{Nm/rad}$.
This value represents the stiffness required for the knee when running at $5\si{m/s}$ \cite{rummel2008stable}
and is reasonable as an upper bound in legged locomotion at $1\si{m/s}$.
The stiffness of FI controller was set to various values ranging 
from $100  ~ \si{Nm/rad}$ to $1800  ~ \si{Nm/rad}$. 
The range of each angle of $\bm{\theta}_{eq}$ is shown in Table. \ref{tab:angle range}.
$c_{kd}$ is set as $c_{kd} = 0.02 \si{N m s/rad}$ in (\ref{eq:k_to_d}).

\begin{table}[H]
	\centering
	 \caption{Range of joint angle $[\si{rad}]$} \label{tab:angle range}
	 \begin{tabular}{c|c|c|c} 
        & $\theta_{eq1}$ & $\theta_{eq2}$ & $\theta_{eq3}$ \\
        \hline
        $\max$ & $2\pi/3$ & $0$ & $\pi/3$ \\
        $\min$ & $-\pi/3$ & $-17 \pi /18$ & $-\pi/3$
	 \end{tabular}
\end{table}

In the simulation, joint constraints are not configured
to prevent the agent from utilizing soft contact constraints of Mujoco \cite{todorov2012mujoco} for locomotion.
For joint actuation setting, 
any models, as friction and armuture are not implemented
rather than the stiffness matrix and damping matrix
coming from each controller settings.
As shown in Fig.\ref{Fig:hierachy_control_diagram}, the ground surface in the physical simulation is uneven.
The height of the ground is randomly determined every 0.02 m in the forward direction,
following a uniform distribution $U(0,0.03)$.

\subsection{Reward Design}
\label{sec:reward_design}


For the monoped locomotion task,
the immediate reward for one time step consists 
of four reward functions,
as
\begin{align}
    r = r_{hlty} + r_{for}  + r_{ctrl} + r_{knee}. \label{eq:monoped_reward}
\end{align}
In addition to $r_{hlty}$, $r_{for}$ and $r_{ctrl}$ used in \textit{Hopper-v4} environment in \textit{Open-AI Gym},
$r_{knee}$ is designed.


$r_{hlty}$ is the reward given when an agent does not fall down in the time step
and designed as
\begin{align}
    r_{hlty} = 
    \left\{
    \begin{array}{ll}
    1     & \begin{array}{l}
            - \pi /12 < \phi_b <  \pi/6   \\ 
          \qquad \qquad \rm{and} \\ 
           0.9\si{m} < z < 2.1\si{m}
    \end{array} \\
    0      &  \rm{otherwise}
    \end{array}
    \right. . \label{eq:terminate}
\end{align}
If $r_{hlty} = 0$, the episode is terminated at the time step.
$r_{for}$ is the reward given depending on the velocity of the center of torso mass as
\begin{align}
    r_{for} = 
    \left\{
    \begin{array}{ll}
    \dot{x}     &  \dot{x} \leq \dot{x}_d \\
    2\dot{x}_d - \dot{x}     & \dot{x} > \dot{x}_d  
    \end{array}
    \right. .
\end{align}
$\dot{x}_d$ is the target velocity
and
is set as $\dot{x}_d = 1 \si{m/s}$.
$r_{ctrl}$ is  the regularization term for torque averaged by the number of control duration steps $N_{step}$ and
\begin{align}
    r_{ctrl} = - 2\times 10 ^{-6} \cdot \frac{1}{N_{step}} \sum_{j=0}^{N_{step-1}} || \bm{\tau} (t_{sim,j}) ||^2  .
\end{align}
From $\nu_c$ and the simulation frequency $\nu_s = 600\si{Hz}$,
$N_{step}$ means the total number of simulation steps included in one control step and is determined as $N_{step} = \nu_s/\nu_c$.
$t_{sim,j} (j \in [0, N_{step}-1] )$ is number of, $j$-th simulation step included in one control step.
Similarly,
$r_{knee}$ is designed as
\begin{align}
     r_{knee} &= - 0.1 \cdot \frac{1}{N_{step}} \sum_{j=0}^{N_{step}-1} f_{knee} (t_{sim,j}) \\
     f_{knee}(t_{sim,j}) &= 
     \left\{
     \begin{array}{ll}
     1     &  \theta_2>0\\
     0     &  \rm{otherwise}
     \end{array}
     \right. 
\end{align}
for preventing back knee locomotion.

 For the biped locomotion task,
 two an additional reward function, $r_{cross}$ and $r_{grd}$ is given.
 To facilitate alternating leg forward gait,
 $r_{cross}$ is given.
 $r_{cross}$ outputs the positive value, 
 when the front-back relationship of the left and right legs is interchanged 
 from one step before.
 It is expressed as
 \begin{align}
     r_{cross} = \left\{ 
     \begin{array}{ll}
     2\times 10 ^{-2} \cdot \nu_c     &  
          \rm{if ~ take ~ a ~ step}\\
     0     &  \rm{otherwise}
     \end{array} \right. .
 \end{align}
 Even if the control frequency is changed, 
 the balance of each reward function should be the same.
 Then, $r_{cross}$ is depending on $\nu_c$.
 To prevent diversity of gait,
 non touching ground penalty is desined as
 \begin{align}
     r_{grd} &= - 0.5 \cdot \frac{1}{N_{step}} \sum_{j=0}^{N_{step}-1} f_{grd} (t_{sim,j}) \\
     f_{grd}(t_{sim,j}) &= 
     \left\{
     \begin{array}{ll}
     1     &  \rm{both ~ leg ~ in ~ flight ~ phase}\\
     0     &  \rm{otherwise}
     \end{array}
     \right.
 \end{align}
 The total reward for the biped locomotion is
 \begin{align}
     r = r_{hlty} + r_{for}  + \left( r^{l}_{ctrl} + r^{r}_{ctrl} \right)  +  r_{cross} + r_{grd}.
 \end{align}
 $r_{ctrl}$ is given for left leg and right leg, respectively.


\subsection{Learning algorithm}
For the learning algorithm of neural networks,
Soft Actor Critic (SAC) was chosen \cite{pmlr-v80-haarnoja18b}
implemented in stable-baselines 3 library \cite{raffin2021stable},
as the standard algorithm in the reinforcement learning region.
SAC pools training data in replay buffer as shape, $\{\bm{o}_t, \bm{a}_t, \bm{o}_{t+a}, r_t\}$.
In this paper,
to get symmetry gait for biped locomotion,
the mirrored training date $\{\bm{o}^{mirror}_t, \bm{a}^{mirror}_t, \bm{a}^{mirror}_{t+1}, r_t\}$ 
was also pooled,
in which the left and right leg were being swapped in observation and action data.

In this study, learning trials are conducted by varying the control frequency,
and the discount rate $\gamma$ 
depends on the control frequency.
By using one control time step $\Delta t$ and a time coefficient $T_{\gamma}$,
$\gamma$ were rescaled based on $\nu_c$ \cite{doya2000reinforcement} as
\begin{align}
    \gamma = 1 - \frac{\Delta t}{\tau}. \label{eq:Doya}
\end{align}
By substituting $\Delta t = 1/\nu_c$ to (\ref{eq:Doya}),
the discount rate is expressed as
\begin{align}
    \gamma = 1 - c_\nu \cdot \frac{1}{\nu_c}
\end{align}
to get the same discount after the same amount of time has passed
for each $\nu_c$.
$c_\nu = 1/\tau$ is a coefficient for the discount rate 
and $c_\nu = 0.5$ was chosen from preliminary experiments.


For each method, training was performed five times under each control frequency condition and each observation condition.
The control frequencies were set to 20, 10, 5, and 3 Hz for the monoped locomotion task,
and 10, 5, and 3 Hz for the biped locomotion task.
Additionally, the maximum trial duration was set to 20 seconds, 
and (\ref{eq:terminate}) was used as the termination condition for falls.

\section{Results}
\label{sec:Results}

\subsection{Monoped locomotion task}

\begin{figure}[tb]
	\centering
    \begin{minipage}{\hsize}
    \centering
        \includegraphics[width=0.6\hsize]{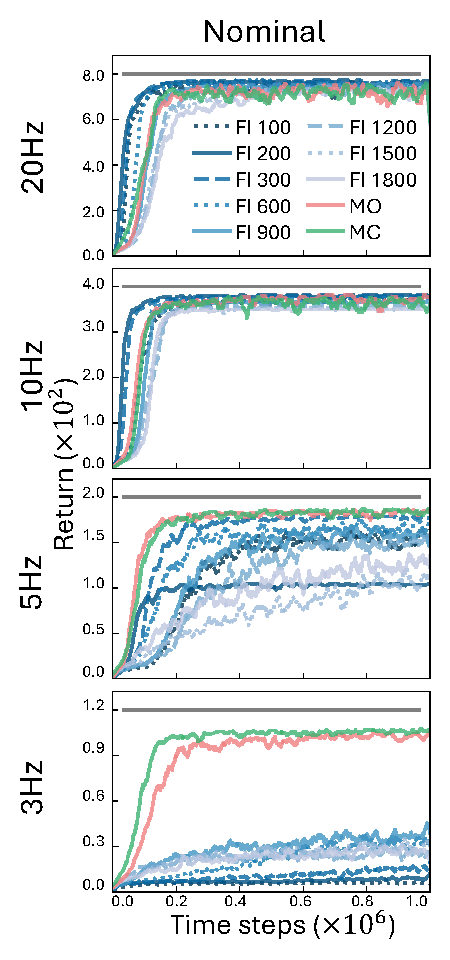}
    \end{minipage}
	\caption{Learning curve of monoped locomotion task under nominal observation condition.
    The slower the control frequency is, the more apparent the advantage of MC and MO becomes.}
        \label{fig:LRcurve_monoped_full}
\end{figure}

\begin{figure*}[tb]
	\centering
    \begin{minipage}{\hsize}
    \centering
        \includegraphics[width=\hsize]{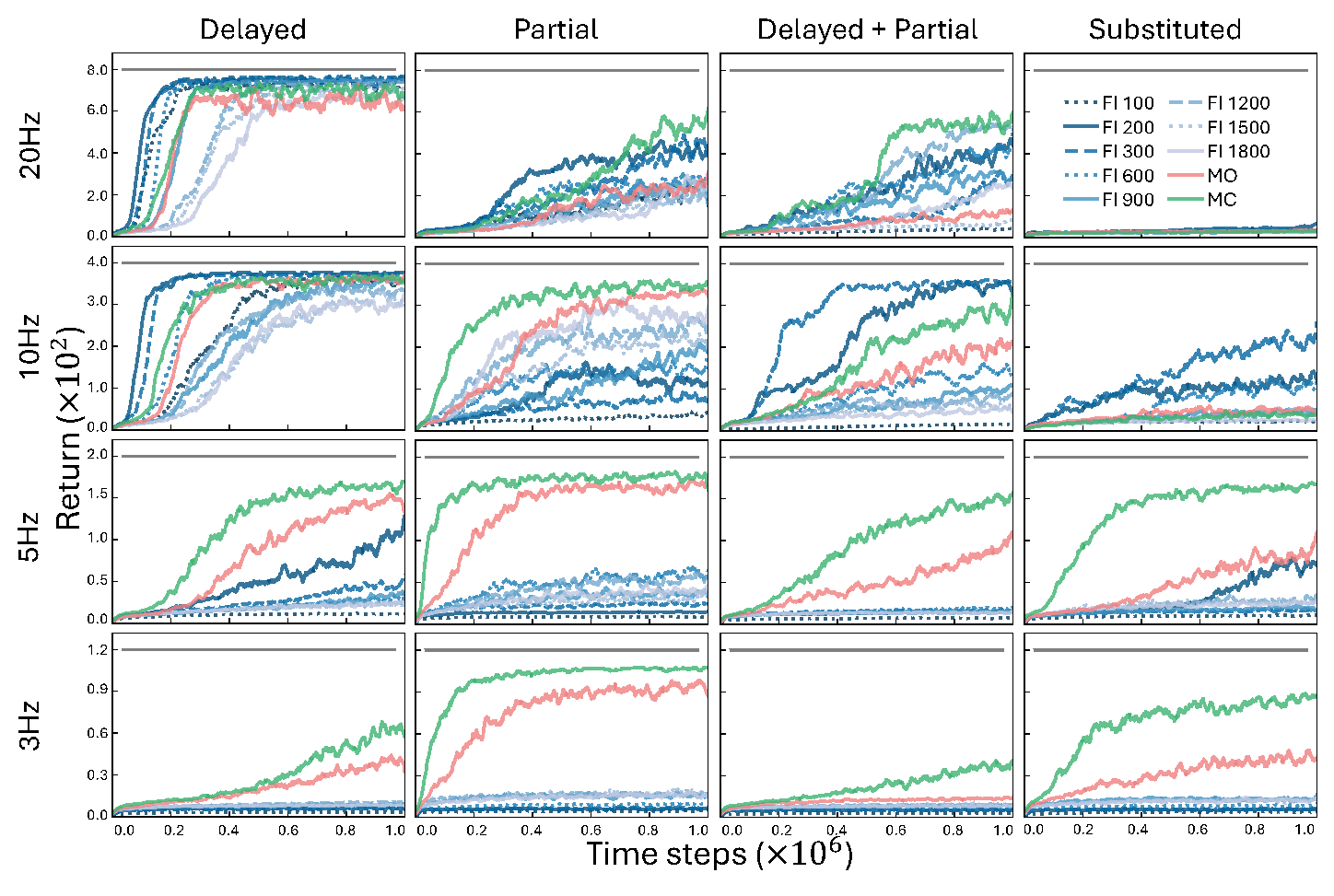}
    \end{minipage}
	\caption{Learning curve of monoped locomotion task under limited observation conditions.}
        \label{fig:LRcurve_monoped}
\end{figure*}


Fig.~\ref{fig:LRcurve_monoped_full} and~\ref{fig:LRcurve_monoped} show the learning curves 
in monoped locomotion task
for nominal 
and the other limited observation, respectively.  
Each curve represents the mean of five independent training trials at each step.  
For visibility, the standard deviation is omitted in this study.  
The black solid line indicates the theoretical upper bound,
corresponding to perfect tracking of the target forward velocity at every timestep.  
For selected conditions, 
the success rates and return scores are summarized in Tables~\ref{tab:success_rate} and~\ref{tab:learning_score}.  
Each controller was evaluated with 100 test episodes for each of the five trained policies, 
and the averages were computed.  
The return scores were normalized by the upper bound corresponding to each control frequency, 
so that the maximum possible value is 1. 
A trial was regarded as successful if the agent maintained a stable gait for 20~seconds without falling, as defined in (\ref{eq:terminate}).  

\begin{figure}[tb]
	\centering
        \begin{minipage}{\hsize}
	    \centering
	    \includegraphics[width=1\hsize ]{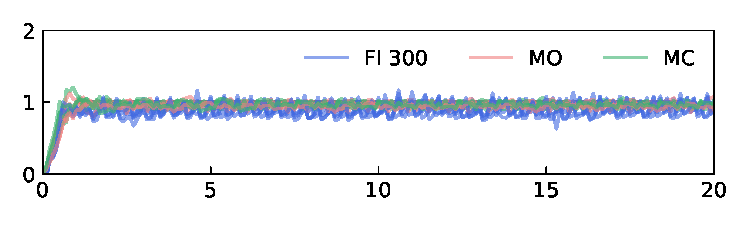}
            \subcaption{5Hz, Nominal observation}
            \label{fig:5_Def}
	\end{minipage}
        \begin{minipage}{\hsize}
	    \centering
	    \includegraphics[width=\hsize]{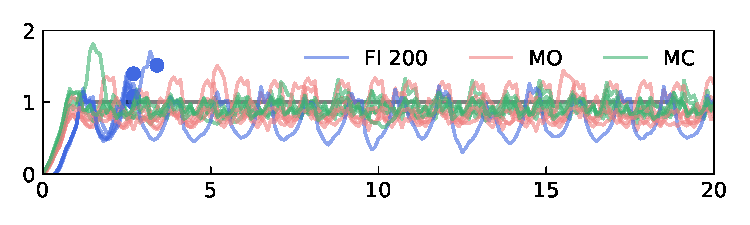}
            \subcaption{3Hz, Nominal observation}
            \label{fig:3_Def}
	\end{minipage}
        \begin{minipage}{\hsize}
	    \centering
	    \includegraphics[width=\hsize]{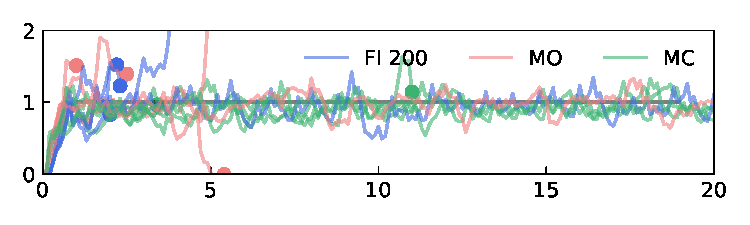}
            \subcaption{5Hz, Substituted observation}
            \label{fig:5_Som}
	\end{minipage}
	\caption{Horizontal velocity trajectories $\dot{x}$ while following the target velocity 
    $\dot{x}_d = 1~\si{m/s}$ in the monoped locomotion task.
                Each dot means the falling down and the end of the trial.
                (a) Almost all agents were converging to steady locomotion.
                (b) All policies of MC and MO achieved tasks,  many policies of FI 200 fell down.
                (c) More MC policies achieved tasks than MO and FI 200.
                  }
        \label{fig:x_dot_following}
\end{figure}

Fig.~\ref{fig:x_dot_following} shows examples of the $\dot{x}$ tracking $\dot{x}_d = 1 \si{m/s}$ 
on a flat floor for selected conditions and controllers.  
Dots indicate a fall, which triggers the termination condition defined in (\ref{eq:terminate}).  
Because the environment is flat and all policies are deterministic,  
the results in Fig.~\ref{fig:x_dot_following} are reproducible and consistent across runs.


Under nominal observation condition (Fig.~\ref{fig:LRcurve_monoped_full}),  
the FI controllers (FI 200, FI 300) learned faster 
than MO and MC at 20~Hz and 10~Hz.  
However, this tendency reversed at slow-rate control frequencies.
At 5~Hz and 3~Hz,  
the morphological computation effect became evident,  
and both MC and MO achieved locomotion while FI controllers failed to learn stable gaits.  
As shown in Tables~\ref{tab:success_rate} and~\ref{tab:learning_score},  
the performance gap among controllers is small at 10~Hz and 5~Hz,  
but at 3~Hz, MC and MO outperform FI significantly.  
Figs.~\ref{fig:5_Def} and~\ref{fig:3_Def} illustrate the forward velocity $\dot{x}$ on level ground at 5~Hz and 3~Hz, respectively.  
All policies succeed at 5~Hz, but only MC and MO maintain locomotion at 3~Hz,  
while FI controllers were likely to fall down shortly after starting.  
Nevertheless, the performance difference between MC and MO remains minor under the nominal observation.


In delayed observation condition (Fig.~\ref{fig:LRcurve_monoped}, first column),  
at 20~Hz and 10~Hz, some FI controllers still learned faster,  
similar to the nominal condition.  
This is because a one-step delay corresponds to a short physical time at high control frequencies.  
However, at 5~Hz, where the delay becomes significant,  
learning stagnates for all FI controllers except FI~200,  
while MC and MO still achieve locomotion.  
The results in Tables~\ref{tab:success_rate} and~\ref{tab:learning_score} confirm  
a clear advantage of MC and MO under delayed observations.


In partial observation condition (Fig.~\ref{fig:LRcurve_monoped}, second column),  
learning curves at 20~Hz and 10~Hz indicate that fast-rate control frequency does not necessarily make the task easier.  
This trend also holds true when angular velocity cannot be observed, 
and similarly under delayed partial as well as substituted observations.
At 5~Hz and 3~Hz, MC and secondarily MO show clearer advantages,
while the all FI controllers failed to learn locomotion,  
as also supported by Tables~\ref{tab:success_rate} and~\ref{tab:learning_score}.  
The difference in learning efficiency is observed between MC and MO,  
suggesting that biarticular coupling provides some benefit under partial observation.


Under delayed partial observation condition (Fig.~\ref{fig:LRcurve_monoped}, third column)  ,
at 10~Hz, some FI controllers perform slightly better than in the partial condition
due to including the previous action as part of the observation vector, observation condition
but at 5~Hz, learning again fails.
Meanwhile, MC was achieved locomotion, just like under the other observation conditions,
as shown in Tables~\ref{tab:success_rate} and~\ref{tab:learning_score}. 
The performance gap between MC and MO becomes more apparent than in delayed observation or partial observation.


Finally, under substituted observation condition (Fig.~\ref{fig:LRcurve_monoped}, fourth column),  
both MC and MO failed to learn at 20~Hz and 10~Hz,  
and none of the FI controllers achieved stable locomotion either.  
At slow-rate control frequencies (5~Hz and 3~Hz), however,  
MC were more likely to achieve locomotion than MO and FI, as confirmed in  
Table~\ref{tab:success_rate}, Table~\ref{tab:learning_score}, and Fig.~\ref{fig:5_Som}.  

These results indicate that the morphological computation inherent in  
the variable-impedance muscle coordination allows the system  
to compensate with limited observation at slow-rate control frequency
in monoped locomotion task.

\begin{table}[htbp]
  \centering
  \scriptsize
  \caption{Success rate ($\%$) in monoped locomotion task}
  \label{tab:success_rate}
  \begin{tabular}{l|l|l|l|l|l|l}
    &&& \multicolumn{4}{c}{FI} \\
    & MC & MO & 200 & 300 & 600 & 900 \\
    \hline 
    $10\si{\hertz}$ Nominal & 96.0 & 98.4 & 99.8 & 99.8 & \textbf{100} &\textbf{100} \\
    \hline 
    $5\si{\hertz}$ Nominal & 97.4 & 99.2 & 59.8 & \textbf{99.6} & 92.3 & 82.0 \\
    \hline 
    $5\si{\hertz}$ Delayed & \textbf{92.4} & 74.8 & 39.6 & 2.8 & 0.0 & 0.4 \\
    \hline 
    $5\si{\hertz}$ Partial & \textbf{95.8} & 91.8 & 0.0 & 0.0 & 8.8 & 0.2 \\
    \hline 
    \begin{tabular}[t]{@{}l@{}}
    $5\si{\hertz}$ Delayed \\ $+$ Partial
    \end{tabular} & \textbf{84.6} & 32.4 & 0.0 & 0.0 & 0.0 & 0.0 \\
    \hline 
    $5\si{\hertz}$ Substituted & \textbf{95.8} & 23.0 & 16.0 & 0.0 & 0.0 & 0.0 \\
    \hline 
    $3\si{\hertz}$ Nominal & 92.2 & \textbf{96.6} & 0.0 & 0.0 & 8.4 & 6.6 \\
    \hline 
    $3\si{\hertz}$ Substituted & \textbf{75.2} & 11.6 & 0.0 & 0.0 & 8.4 & 6.6 \\
  \end{tabular}
\end{table}

\begin{table}[htbp]
  \centering
  \scriptsize
  \caption{Averaged return in monoped locomotion task. 
           The each value is normalized by the upper bound. }
  \label{tab:learning_score}
  \resizebox{\linewidth}{!}{
  \begin{tabular}{l|l|l|l|l|l|l}
    &&& \multicolumn{4}{c}{FI} \\
    & MC & MO & 200 & 300 & 600 & 900 \\
    \hline 
    $10\si{\hertz}$ Nominal & 0.912 & 0.943 & \textbf{0.952} & 0.949 & 0.936 & 0.914 \\
                             & $\pm$0.181 & $\pm$0.101 & $\pm$0.041 & $\pm$0.041 & $\pm$0.004 & $\pm$0.006 \\
    \hline 
    $5\si{\hertz}$ Nominal & 0.914 & \textbf{0.924} & 0.519 & 0.891 & 0.856 & 0.782 \\
                             & $\pm$0.133 & $\pm$0.081 & $\pm$0.414 & $\pm$0.019 & $\pm$0.162 & $\pm$0.220 \\
    \hline 
    $5\si{\hertz}$ Delayed & \textbf{0.818} & 0.754 & 0.585 & 0.247 & 0.154 & 0.172 \\
                             & $\pm$0.149 & $\pm$0.226 & $\pm$0.280 & $\pm$0.172 & $\pm$0.080 & $\pm$0.140 \\
    \hline 
    $5\si{\hertz}$ Partial & \textbf{0.886} & 0.846 & 0.072 & 0.120 & 0.273 & 0.203 \\
                             & $\pm$0.150 & $\pm$0.125 & $\pm$0.024 & $\pm$0.075 & $\pm$0.255 & $\pm$0.147 \\
    \hline 
    $5\si{\hertz}$ Delayed & \textbf{0.768} & 0.505 & 0.074 & 0.081 & 0.090 & 0.080 \\
            $+$ Partial      & $\pm$0.174 & $\pm$0.278 & $\pm$0.021 & $\pm$0.020 & $\pm$0.043 & $\pm$0.027 \\
    \hline 
    $5\si{\hertz}$ Substituted & \textbf{0.832} & 0.429 & 0.341 & 0.081 & 0.087 & 0.098 \\
                             & $\pm$0.105 & $\pm$0.304 & $\pm$0.280 & $\pm$0.031 & $\pm$0.044 & $\pm$0.052 \\
    \hline 
    $3\si{\hertz}$ Nominal & \textbf{0.884} & 0.862 & 0.073 & 0.117 & 0.297 & 0.312 \\
                             & $\pm$0.092 & $\pm$0.082 & $\pm$0.047 & $\pm$0.107 & $\pm$0.262 & $\pm$0.240 \\
    \hline 
    $3\si{\hertz}$ Substituted & \textbf{0.724} & 0.362 & 0.049 & 0.050 & 0.075 & 0.118 \\
                             & $\pm$0.197 & $\pm$0.244 & $\pm$0.022 & $\pm$0.023 & $\pm$0.031 & $\pm$0.053 \\
  \end{tabular}
  }
\end{table}

\subsection{Biped locomotion task}

\begin{figure}[tb]
	\centering
    \begin{minipage}{\hsize}
    \centering
        \includegraphics[width=\hsize]{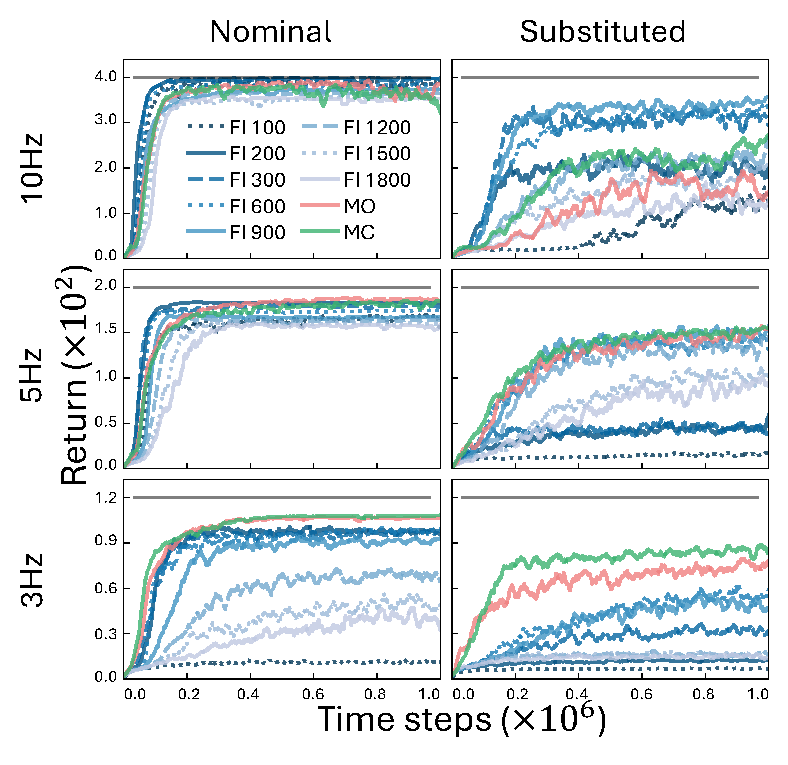}
    \end{minipage}
	\caption{Learning curve of monoped locomotion task under nominal and substituted observation conditions.}
        \label{fig:LRcurve_biped}
\end{figure}

Fig.\ref{fig:LRcurve_biped} shows the learning results of the biped locomotion task 
under each experimental condition.  
The black solid line represents the reference bound corresponding to perfect tracking of the target velocity at each timestep.  
Note that this is not a strict upper bound due to the inclusion of the $r_{cross}$ term in the reward function.

Under the nominal observation condition (Fig.~\ref{fig:LRcurve_biped}, first column),
in contrast to the monoped locomotion results,
stable learning was achieved not only with the MC and MO controllers but also with appropriately tuned FI controllers 
even at 3~Hz.  
This is attributed to the inherent stability and redundancy of the biped locomotion task, 
which makes it easier to achieve compared to the monoped case.  
However, when compared across control frequencies (10~Hz and 5~Hz),
the MC and MO controllers converged faster and achieved higher return values.  
Overall, these results resemble those of the monoped locomotion task under the nominal observation condition from 20~Hz up to 5~Hz (Fig.~\ref{fig:LRcurve_monoped_full}).  
For the main successful controllers, 
a stable alternating gait pattern, where both legs take turns stepping forward, was observed.

Under the substituted observation condition (Fig.~\ref{fig:LRcurve_biped}, second column),
the advantages of MC and MO become apparent at 5~Hz and 3~Hz.  
However, under this condition, 
the learned gaits varied considerably even within the same controller settings.  
The agents often developed asymmetric, cross-free gaits, 
moving their legs in a fore–aft manner similar to the front and hind limbs in quadrupedal locomotion
as well as symmetric alternating walking gaits.  
As a result, the learning curves exhibited greater variability across trials.




Although less evident than in the monoped task, 
the biped results still indicate that variable-impedance muscle coordination 
supports locomotion under slow-rate control frequency and limited-observation condition.


\section{Discussion}
\label{sec:Discussion}

\subsection{Morphological computation by muscle coordination} \label{sec:morphological}

We conduct our analysis on monoped locomotion.
In the MC controller, the high-level neural network must output not only the equilibrium joint angles but also the variable stiffness values for each muscle-coordination element. This requirement increases both the number of parameters and the dimensionality of the search space, potentially making the learning process more complex and slower.
Thus, although low-level morphological computation can facilitate stable control, there may be a trade-off between its benefits and the increased learning cost introduced by the higher-dimensional parameter search.

Indeed, as shown in Fig.~\ref{fig:LRcurve_monoped_full},
under the nominal observation condition, some FI controllers converged faster than MC and MO at fast-rate control frequencies (20~Hz and 10~Hz).
In contrast, MC and MO clearly outperformed FI at slow-rate control frequencies (5~Hz and 3~Hz).
These results suggest that while the advantages of morphological computation in muscle coordination are less pronounced at high control frequencies, they become evident under slow-rate control conditions, where the feedback capability of the high-level controller is more limited.



In addition, as shown in Fig.~\ref{fig:LRcurve_monoped},
the results obtained under other limited-observation conditions at slow-rate control frequencies indicate that the morphological computation provided by muscle coordination effectively reshapes the sensor–motor mapping from the perspective of the high-level controller.


Specifically,
(a) delays became easier for the high-level controller to infer and compensate for,
(b) fast dynamics represented by angular velocities no longer needed to be directly observed, and
(c) actual joint angles could be replaced by equilibrium angles as an efference copy.
These characteristics align with fundamental properties observed in human and biological motor control, suggesting that such capabilities are inherently embedded within the muscle-coordination mechanism.
The implication of aspect (c) is further elaborated in Sec.~\ref{sec:self-organ}.

\subsection{Self-organization and muscle coordination} \label{sec:self-organ}



This section examines the self-organization that becomes particularly evident in the MC controller under the substituted-observation condition in the monoped locomotion task.
Compared with the nominal condition, where the full $11$-dimensional external state is observable, only 4 dimensions were available in this constrained setting.
Despite this significant reduction, the MC controller achieved stable locomotion, suggesting that the effective dimensionality required for control was reduced through self-organization within the low-level dynamics, including the MC controller itself.
More concretely, a self-organized relationship emerged between the actual joint angles $\bm{\theta}$ and the equilibrium angles $\bm{\theta}_{eq}$.

\begin{figure}[tb]
	\centering
        \includegraphics[width=\hsize]{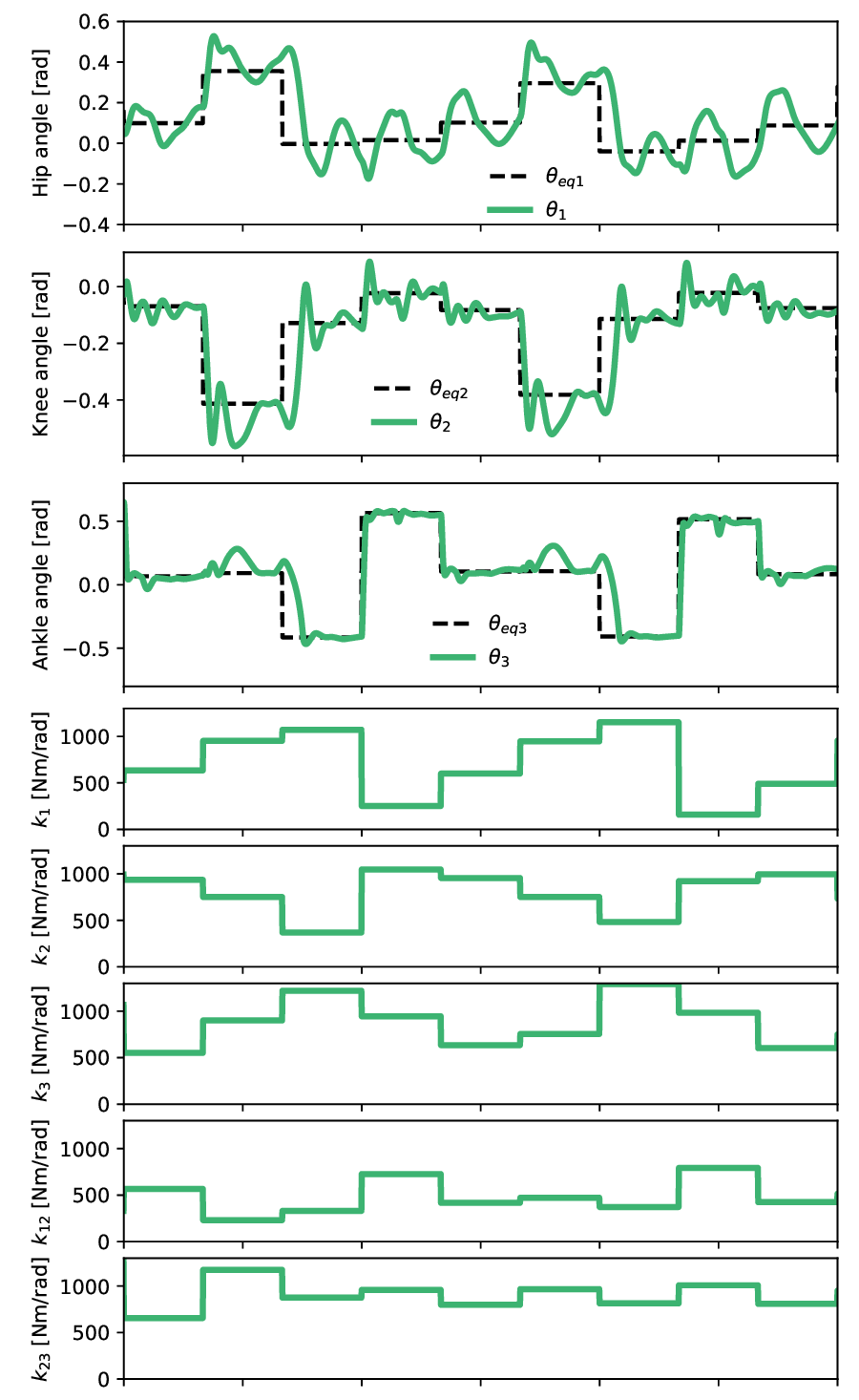}
	\caption{Time series of leg angles and stiffness inputs in the monoped locomotion task at $3$ Hz under the substituted-observation condition with the MC controller.
Each leg angle converges toward its corresponding equilibrium angle, demonstrating that the controller can effectively utilize equilibrium angles in place of actual joint-angle observations for stable locomotion.}
        \label{fig:angle_follow_angle_eq}
\end{figure}

Fig.~\ref{fig:angle_follow_angle_eq} presents the time series of leg angles and stiffness inputs during the monoped locomotion task at $5$ Hz under the substituted observation condition with the MC controller.
As shown, $\bm{\theta}$ closely tracks $\bm{\theta}{\mathrm{eq}}$, and although their difference is not zero at every control step, it exhibits a clear convergent pattern.
This suggests that the neural network effectively used the previous equilibrium-angle action $\bm{a}{\theta}^{t-1}$ as a surrogate for the actual joint angles $\bm{\theta}^{t}$,
indicating that a convergent dynamical relationship between $\bm{\theta}^{t}$ and $\bm{\theta}_{\mathrm{eq}}^{,t-1}$ emerged spontaneously.
Such self-organized dynamics were particularly pronounced in the MC controller, which incorporates variable biarticular impedance.


This result suggests that the system generated motion consistent with the virtual trajectory hypothesis, as the actual joint angles $\bm{\theta}$ closely followed the equilibrium trajectory $\bm{\theta}{\mathrm{eq}}$~\cite{bizzi1984posture}.
At the same time, because not only $\bm{\theta}{\mathrm{eq}}$ but also the stiffness $\bm{K}$ varied over time, as shown in Fig.~\ref{fig:angle_follow_angle_eq}, the system can also be interpreted as implicitly solving an inverse-dynamics problem. Specifically, it produced appropriate joint torques $\bm{\tau}$ through an internal model~\cite{kawato1999internal} instantiated by the variable-impedance muscle coordination.
Taken together, these findings provide empirical evidence that impedance control (equilibrium-point control) and internal-model-based control can coexist and mutually complement one another within a unified motor-control framework~\cite{burdet2001central,burdet2006stability}.




\subsection{Difference of locomotion style} \label{sec:loco_style}

\begin{figure}[tb]
        \centering
        \includegraphics[width=\hsize]{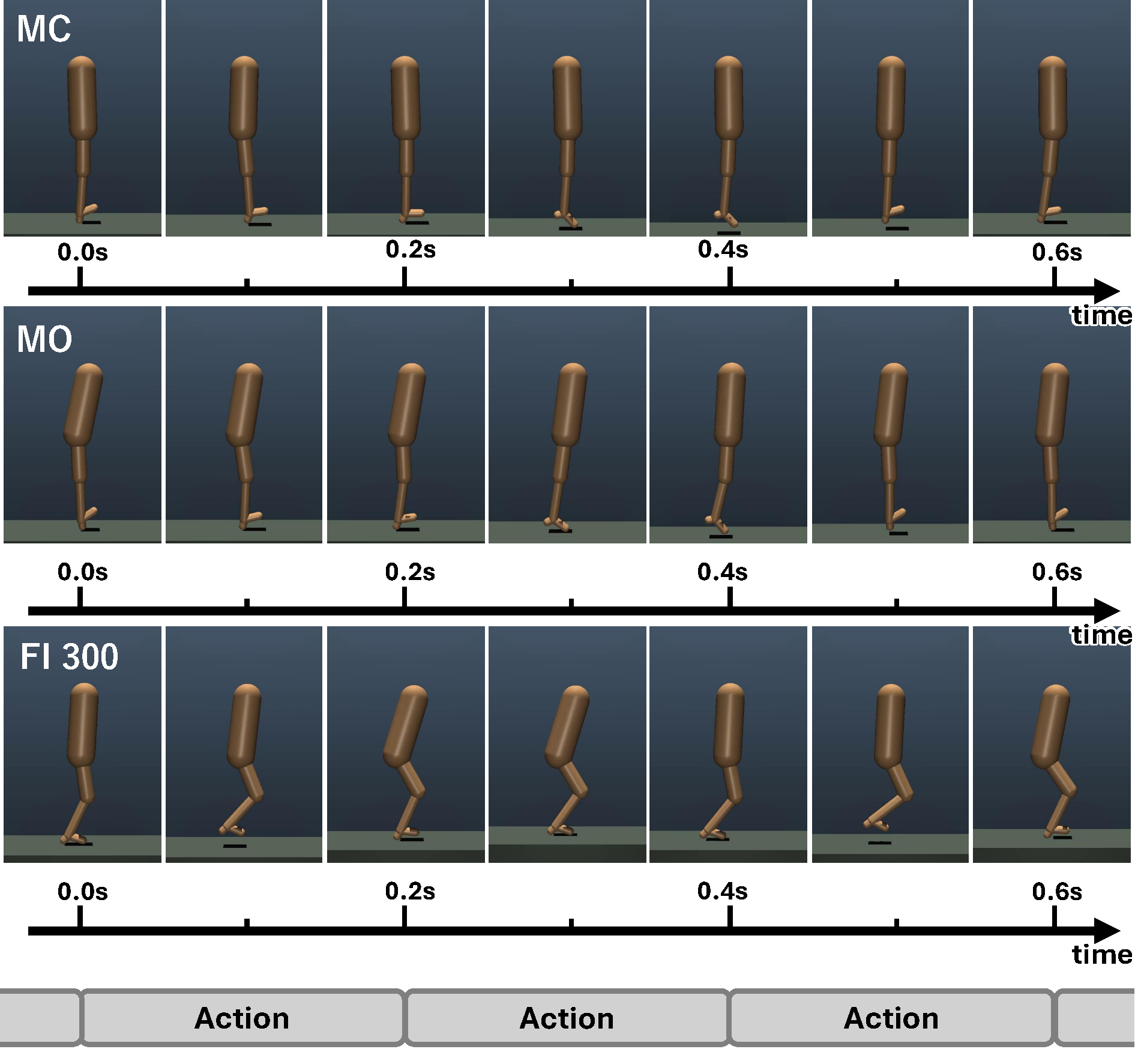}
      \caption{Snapshot of monoped locomotion at 5$\si{Hz}$ under nominal observation.}
      \label{fig:snapshot_monoped_5Hz}
\end{figure}

\begin{figure}[tb]
         \centering
         \includegraphics[width=\hsize]{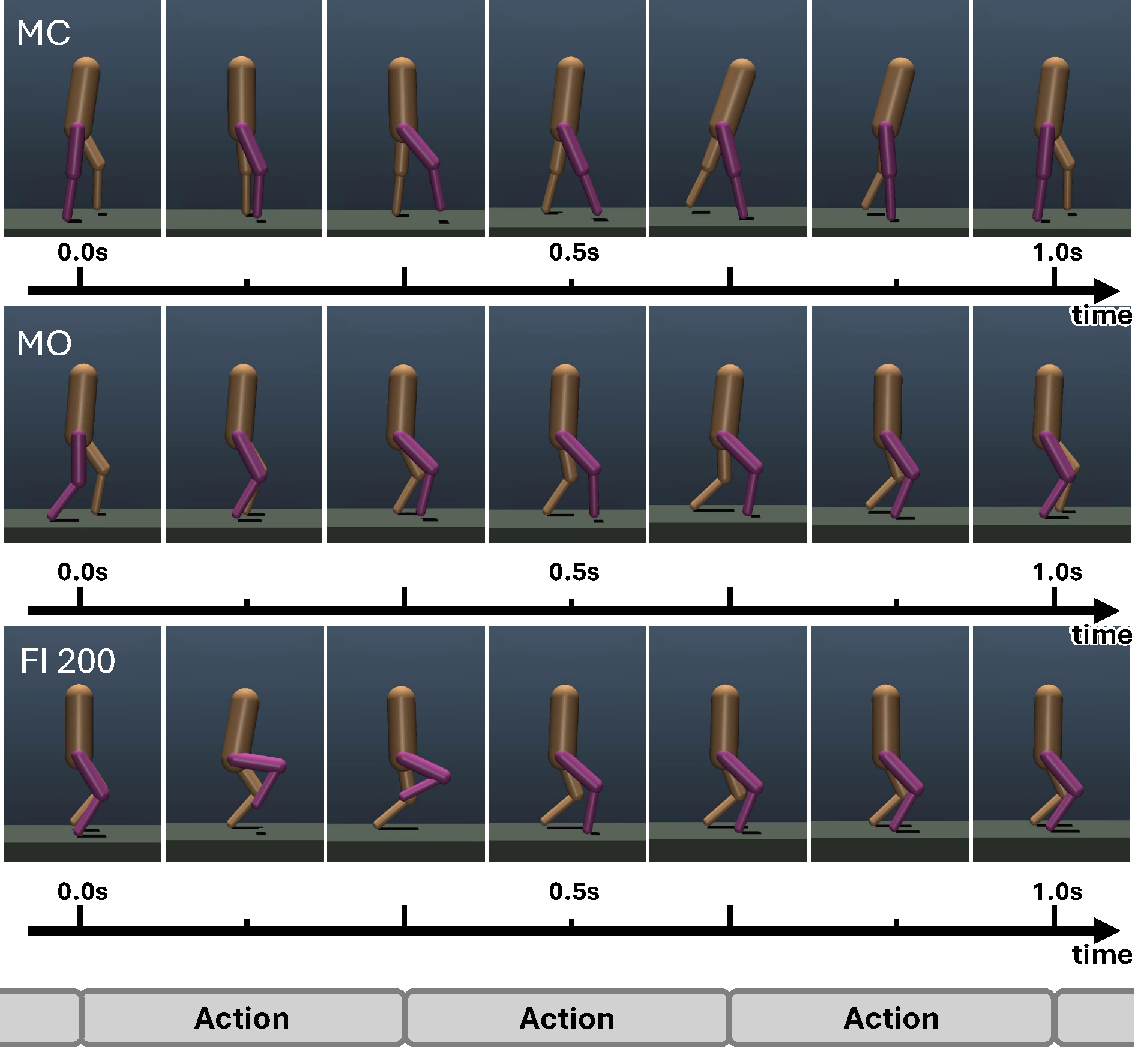}
       \caption{Snapshot of monoped locomotion at 3$\si{Hz}$ under nominal observation}
       \label{fig:snapshot_biped_3Hz}
 \end{figure}

Fig.~\ref{fig:snapshot_monoped_5Hz} shows representative snapshots of monoped locomotion under the nominal observation condition at $5$ Hz for the MC, MO, and FI300 controllers.
Likewise, Fig.~\ref{fig:snapshot_biped_3Hz} presents snapshots of biped locomotion under the nominal observation condition at $3$ Hz for the MC, MO, and FI200 controllers.
The gray shaded regions denote the control periods, during which each controller holds its action output constant.




As the controller type progresses from FI to MO to MC, the knee posture becomes increasingly extended, resulting in a straighter-legged gait in both monoped and biped locomotion. This tendency is particularly pronounced in the biped case.
Such gait characteristics were consistently observed under slow-rate control frequency. 
Given the structure of the reward function in (\ref{eq:monoped_reward}), a straighter-legged gait is advantageous because it reduces the required joint torques, thereby improving the $r_{\mathrm{ctrl}}$ term.
However, excessive knee extension is discouraged by the $r_{\mathrm{knee}}$ penalty, which prevents hyperextension.


This implies that excessive knee extension should be avoided when it compromises stable control.
Therefore, the incorporation of biarticular muscles and variable impedance in the MC controller likely contributes to the emergence of dynamically human-like locomotion patterns that balance torque efficiency with postural stability.

\section{Conclusion}
\label{sec:conclusion}




This study demonstrated that the morphological computation inherent in muscle coordination  
reduces the control requirements of the high-level controller in dynamic motion tasks.  
A hierarchical controller was implemented, consisting of a high-level neural network  
and a low-level variable-impedance muscle coordination,  
and its learning performance was evaluated across various conditions in legged locomotion, mainly monoped locomotion task.
The results showed that even under slow-rate control and biologically inspired observation constraints  
(such as delayed, partial, or substituted observations),  
the variable-impedance muscle coordination achieved stable locomotion,  
revealing clear evidence of morphological computation.  
This indicates that muscle coordination effectively offloads feedback control, 
reduces the bandwidth, and observability required at the high level controller.




These findings provide new insights into bio-inspired hierarchical control frameworks,  
where low-level embodied intelligence can be efficiently utilized to simplify high-level control.  
They highlight the importance of distinguishing the functional roles between high-level and low-level controllers,  
for example in terms of control frequency and observation dependency,  
to fully exploit the morphological computation capabilities. 
This perspective offers design implications for both the mechanical structure of low-level controllers  
and the algorithmic architecture of high-level controllers.


%

\bibliographystyle{IEEEtran}
\bibliography{bib}



\section{Biography Section}
 
\vspace{11pt}



\begin{IEEEbiography}[{\includegraphics[width=1in,height=1.25in,clip,keepaspectratio]{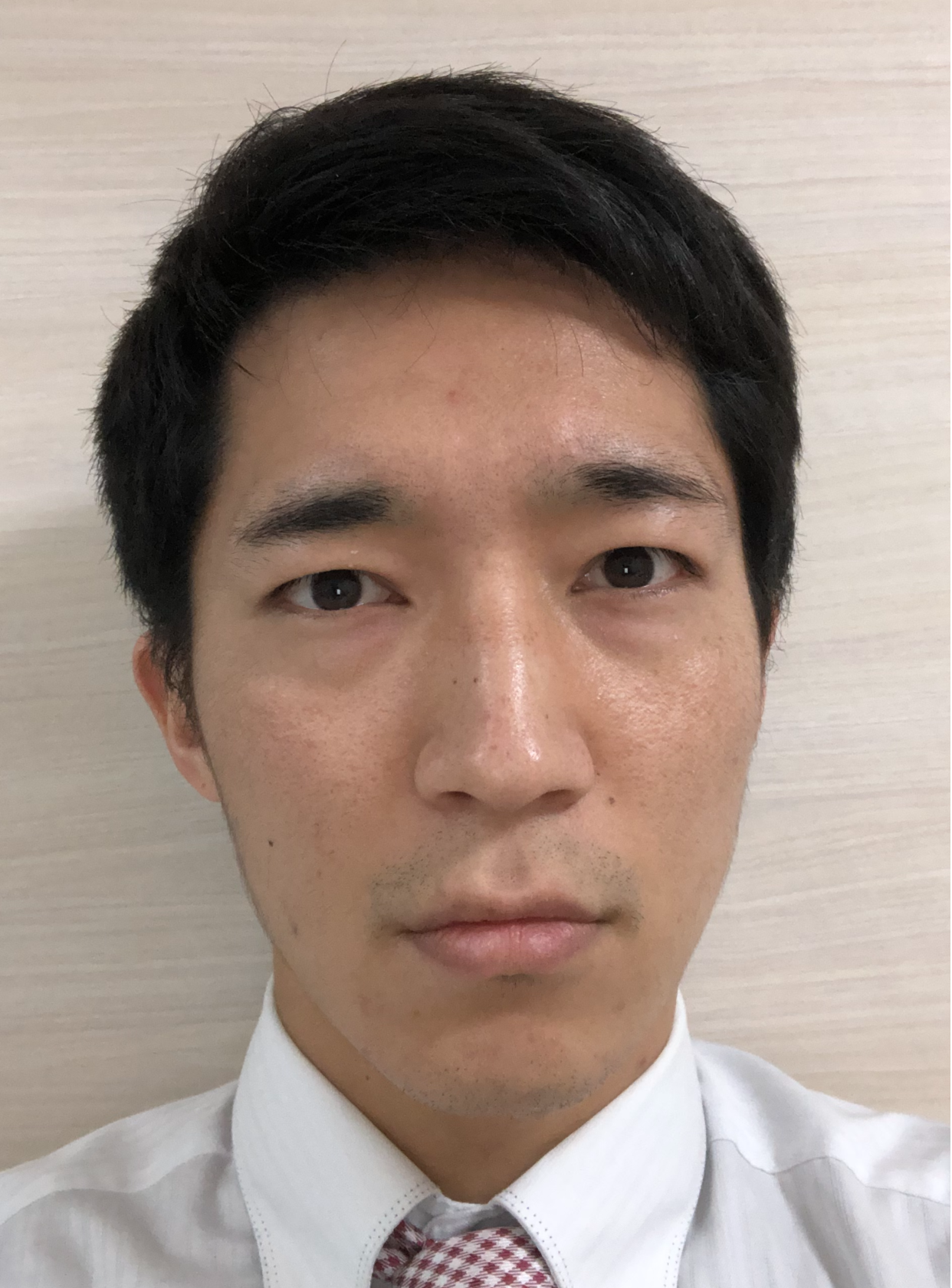}}]{Hidaka Asai} 
received the 
BS degree from the Department of Engineering Science, Kyoto university in 2019,
and the MS degree from
the Kyoto University Graduate school of informatics in 2021.
Now he is a trainee researcher in ATR Computational Neuroscience
Laboratories.
\end{IEEEbiography}

\begin{IEEEbiography}
[{\includegraphics[width=1in,height=1.25in,clip,
trim=100 200 100 200,
keepaspectratio]{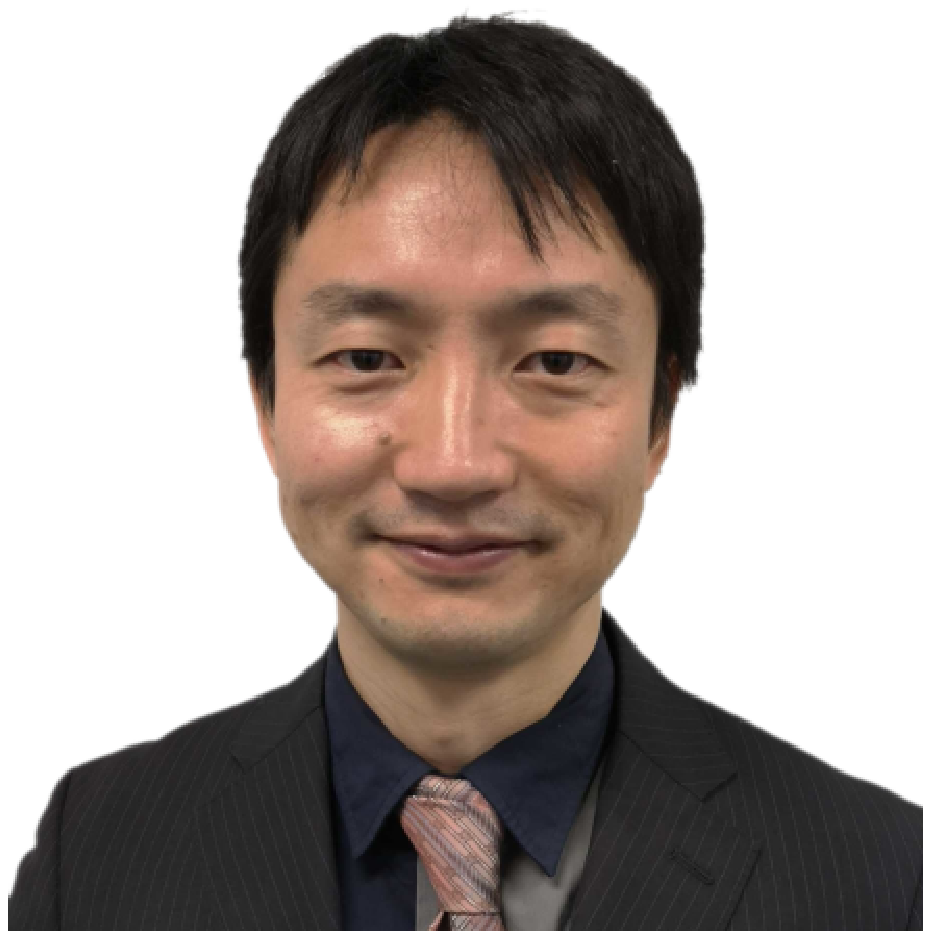}}]{Tomoyuki Noda}
received B.E., M.E., and Ph.D. degrees in Engineering from Osaka University, Osaka, Japan, in 2004, 2006, and 2009, respectively.
He was a JSPS Research Fellow (DC2, 2008--2010). In 2009, he joined the Institute for Neural Computation, University California San Diego, San Diego, CA, USA, as a Visiting Research Scholar. In 2010, he joined to the ATR Computational Neuroscience Laboratories, Kyoto, Japan.
Dr. Noda received the Best Video Nominees Award in the IEEE AAAI Conference on Artificial Intelligence 2008, the Best Video in LAB-RS 2008 for the development of a whole-body humanoid robot with tactile sensation and compliant joints, a Best Paper Nominees Award at IEEE Humanoids 2012, and the BCI Research Award 2017 Top 12 Nominees in a BMI controlled exoskeleton.
\end{IEEEbiography}

\begin{IEEEbiography}
[{\includegraphics[width=1in,height=1.25in,clip,keepaspectratio]{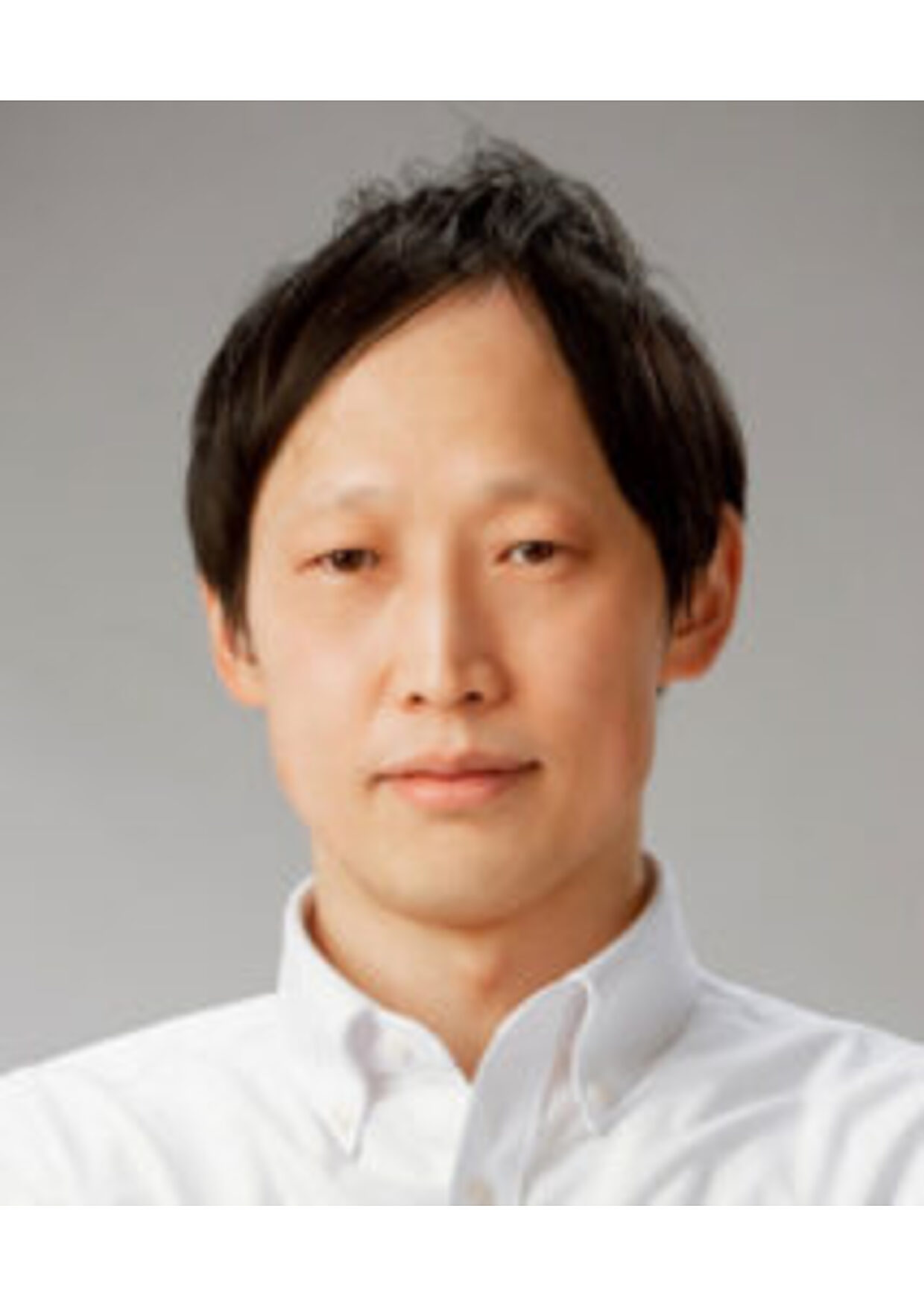}}]{Jun Morimoto}
received the Ph.D.
degree in information science from the Nara Institute
of Science and Technology, Nara, Japan, in 2001.
From 2001 to 2002, he was a Post-Doctoral Fellow with The Robotics Institute, Carnegie Mellon
University, Pittsburgh, PA, USA. He joined the
Advanced Telecommunications Research Institute
International (ATR), Kyoto, Japan, in 2002. He is currently a Professor
with the Graduate School of Informatics, Kyoto University, Kyoto. He is
also the Head of the Department of Brain Robot Interface (BRI), ATR
Computational Neuroscience Laboratories, and a Senior Visiting Scientist
of the Man-Machine Collaboration Research Team, Guardian Robot Project,
RIKEN, Kyoto
\end{IEEEbiography}


\vfill

\end{CJK*}

\end{document}